# Modulation Coding for Flash Memories

Yongjune Kim,
B. V. K. Vijaya Kumar
Data Storage Systems Center
Carnegie Mellon University
Pittsburgh, USA
yongjunekim@cmu.edu

Kyoung Lae Cho, Hongrak Son,
Jaehong Kim, Jun Jin Kong
Flash Development Team
Samsung Electronics Company
Hwasung, Korea
kyounglae.cho@samsung.com

Jaejin Lee
School of Electronic Engineering
Soongsil University
Seoul, Korea
zlee@ssu.ac.kr

*Abstract*—The aggressive scaling down of flash memories has threatened data reliability since the scaling down of cell sizes gives rise to more serious degradation mechanisms such as cell-to-cell interference and lateral charge spreading. The effect of these mechanisms has pattern dependency and some data patterns are more vulnerable than other ones. In this paper, we will categorize data patterns taking into account degradation mechanisms and pattern dependency. In addition, we propose several modulation coding schemes to improve the data reliability by transforming original vulnerable data patterns into more robust ones.

## I. Introduction

In order to meet the market demand for high-density and low-cost flash memories, scaling down and multi-level cell (MLC) together have doubled the memory density every year. However, the aggressive scaling down has given rise to degradation mechanisms such as cell-to-cell interference, program disturb, charge leakage, and lateral charge spreading. These degradation mechanisms reduce the reliability of stored data [1]–[6].

In order to cope with this reliability problem, device level approaches such as improved cell structures and new materials have been considered [3], [5]. In addition, several program schemes have been proposed to reduce the effect of degradation mechanisms [7]–[9]. Another approach is to use stronger error control coding (ECC) and signal processing [10]–[12].

Recently, constrained coding has been investigated to improve the data reliability of flash memories and phase change memories [13]–[16]. The idea is based on the fact that some data patterns are more vulnerable to physical degradation mechanisms than other data patterns. This paper extends such previous works.

In order to deal with these data patterns, we will categorize the vulnerable data patterns and define these vulnerable patterns as E-PH (i.e., the erase state - the highest program state) patterns. In addition, we will explain why these patterns are vulnerable to degradation mechanisms such as cell-to-cell interference and lateral charge spreading. The cell-to-cell interference has been mainly explained by parasitic capacitance between adjacent cells [1], [2]. However, recent research shows that the cell-to-cell interference by direct field effect can become more significant as the cells are scaled down [3], [4]. We will address both the interference due to parasitic capacitance and the interference due to direct field effect.

After characterizing E-PH patterns, we will propose modulation coding schemes to reduce these vulnerable E-PH patterns for single-level cell (SLC), 2-bit/cell and 3-bit/cell MLC. In SLC, the binary run length limited (RLL) codes and non-return to zero inverted (NRZI) coding can reduce the effect of E-PH patterns. In MLC, the conventional binary RLL code can be used to reduce E-PH patterns taking into account the Gray mapping. In addition, we propose novel quaternary and 8-ary constrained codes with higher code rates for 2-bit/cell and 3-bit/cell MLC.

The rest of this paper is organized as follows. In Section II, we explain E-PH patterns considering cell-to-cell interference and lateral charge spreading. In Section III, several modulation coding schemes and their capacities will be presented. In Section IV, via simulations, we will compare the use of modulation coding to the conventional scheme not using modulation coding. Finally, we present our conclusions in Section V.

## II. E-PH Pattern

In $M$-bit/cell flash memory, the cell threshold voltage distribution has $2^M$ states from $S_0$ (the erase state denoted as E) to $S_{2^M-1}$ (the highest program state denoted as PH) as shown in Fig. 1. Each state generally has a binary representation given by Gray mapping. One example of Gray mapping for 2-bit/cell MLC is shown in Table I. Note that the two bits corresponding to a single cell correspond to two different pages.

The E-PH pattern occurs when a cell in E is adjacent to cells in PH. As the number of E-PH patterns increases, the threshold voltage distribution is severely distorted by cell-to-cell interference and lateral charge spreading.

### A. Cell-to-Cell Interference

In flash memory, the threshold voltage shift of one cell can affect the threshold voltage of its adjacent cell. This phenomenon is called cell-to-cell interference (or intercell interference), which is attributed to parasitic capacitance coupling effect between adjacent cells [1], [2].

Fig. 2 shows the conventional flash memory architecture and the cell-to-cell interference. $V_{(i,j)}$ is the threshold voltage of $(i,j)$ cell which is situated at $i$-th word line (WL) and $j$-th bit line (BL). $\gamma_x$ is $x$-directional coupling ratio between BL and adjacent BL. Also, $\gamma_y$ is $y$-directional coupling ratio between WL and adjacent WL. Finally, $\gamma_{xy}$ is $xy$-directional (diagonal) coupling ratio. These coupling ratios depend on parasitic capacitance between adjacent cells. As the cell size continues to shrink, the distances between cells become smaller, which results in the increase of the parasitic capacitance between adjacent cells. The increase of parasitic capacitance causes the increase of coupling ratios [1], [2]. In practice, only the cell-to-cell interference from immediately adjacent cells is considered since the parasitic capacitance diminishes rapidly as the cell-to-cell distance increases [2], [12].

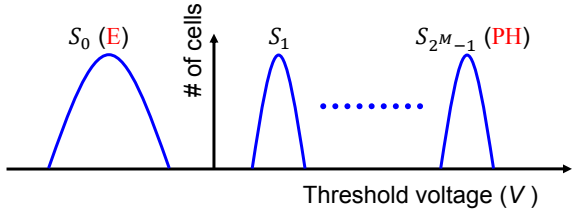

Fig. 1. Threshold voltage distribution of $M$-bit/cell flash memory.

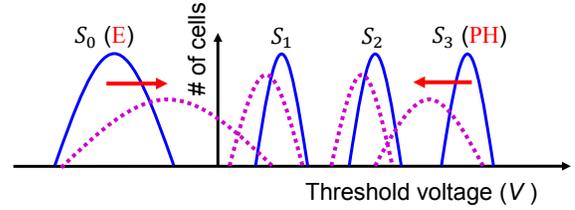

Fig. 3. The distortion of threshold volatage distributions by lateral charge spreading for 2-bit/cell. Solid blue indicates the original distributions and dashed magenta indicates the distributions after lateral charge spreading.

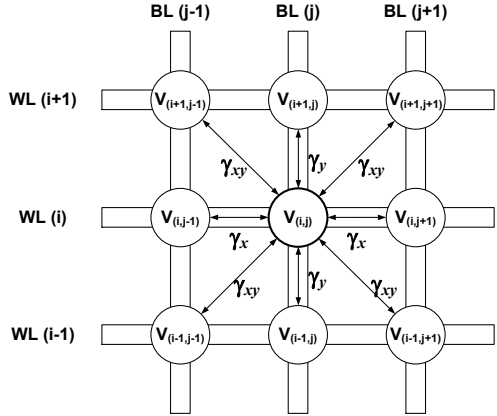

Fig. 2. Conventaion flash memory architecture and cell-to-cell interference.

TABLE I. GRAY MAPPING OF 2-BIT/CELL MLC

|        | $S_0$(E) | $S_1$ | $S_2$ | $S_3$(PH) |
|--------|----------|-------|-------|-----------|
| Page 1 | 1        | 1     | 0     | 0         |
| Page 2 | 1        | 0     | 0     | 1         |

According to [2], the threshold voltage shift $\Delta V_{(i,j)}$ of $(i,j)$ cell due to the cell-to-cell interference can be given by

$$\Delta V_{(i,j)} = \gamma_x \big(\Delta V_{(i,j-1)} + \Delta V_{(i,j+1)}\big) + \\ \gamma_y \big(\Delta V_{(i-1,j)} + \Delta V_{(i+1,j)}\big) + \\ \gamma_{xy} \big(\Delta V_{(i-1,j-1)} + \Delta V_{(i-1,j+1)} + \\ \Delta V_{(i+1,j-1)} + \Delta V_{(i+1,j+1)}\big) \quad (1)$$

which shows that $\Delta V_{(i,j)}$ depends on coupling ratios, the threshold voltage shifts of adjacent cells, and the number of adjacent cells whose threshold voltage shifts are not zero. In literature, the $(i,j)$ cell has been called the victim cell and adjacent cells have been called the aggressor cells.

In order to combat the cell-to-cell interference, various solutions have been proposed. Modifying cell structures and finding novel materials are device level approaches which aim to reduce the coupling ratios [3], [5].

In addition, several program schemes have been proposed to reduce the cell-to-cell interference. These schemes are mainly related to threshold voltage shifts of aggressor cells of (1). The basic idea of these schemes is that a victim cell can be only interfered by its aggressor cells which are programmed after the victim cell has been programmed. Based on this, several program schemes compensate the distortion due to cell-to-cell interference while programming the victim cell [8], [9]. Using these program schemes, the cell-to-cell interference from aggressor cells in the $(i-1)$-st WL can be almost zero and the interference from other aggressor cells can also be significantly reduced.

However, these program schemes are only effective when the victim cell is in program states from $S_1$ to $S_{2^M-1}$. If the victim cell is in E which means that the victim cell is never programmed, these proposed program schemes are not effective since there is no chance to compensate for the cell-to-cell interference from aggressor cells. In addition, when an aggressor cell is at PH, the threshold voltage shift of the aggressor cell is maximized. Therefore, E-PH pattern is the worst pattern when considering the cell-to-cell interference.

Recently, a new cell-to-cell interference mechanism has been discovered. This cell-to-cell interference results from the direct electric field between adjacent cells and is called to direct cell-to-cell interference or channel coupling. The conventional cell-to-cell interference due to parasitic capacitance is sometimes called the indirect cell-to-cell interference in order to distinguish the two cell-to-cell interference mechanisms [3], [4].

As the cell size is scaled down, the direct cell-to-cell interference is predicted to be more serious than the conventional indirect cell-to-cell interference [3]. Particularly, $x$-directional direct cell-to-cell interference is more dominant than interference in other directions [3], [4]. Therefore, $\gamma_x$ of (1) can be changed into $\gamma_x^*$ as follows [3].

$$\gamma_x^* = \gamma_x + \beta \quad (2)$$

where $\gamma_x$ represents coupling ratio by the indirect cell-to-cell interference and $\beta$ represents coupling ratio due to the direct cell-to-cell interference.

*B. Lateral Charge Spreading*

Among the device level attempts for reducing the cell-to-cell interference, charge trap flash (CTF) memories have been proposed as an alternative technology to replace the floating gate (FG) flash memory [5], [6].

However, CTF memories suffer from lateral charge spreading instead of cell-to-cell interference. The lateral charge spreading results from the electric field between adjacent cells. This electric field will be maximized when a cell in E is adjacent to cells in PH. Due to the electric field between adjacent cells, electron charges of a cell spread to other adjacent cells along trap layers [6]. Therefore, cells in E go up to higher states and cells of PH go down to lower states as illustrated in Fig. 3.

In the lateral charge spreading, $x$-directional spreading is much more serious than spreading in other directions. The reason is that trap layers between word lines can be physically separated whereas it is difficult to separate trap layers between bit lines. Therefore, electron charges can move along $x$-directional trap layer more readily [6].

*C. E-PH Pattern*

As stated above, the E-PH pattern is the worst pattern because of cell-to-cell interference and lateral charge spreading. Thus, the number of aggressor cells in PH adjacent to a victim cell in E is important.

There are $256(=2^8)$ E-PH patterns since there are eight aggressor cells for each victim cell. The number of cases of E-PH patterns can be reduced by considering directions since the effect of physical mechanisms such as cell-to-cell interference and lateral charge spreading is generally symmetric in directions. We will classify E-PH pattern as follows.

**Definition 1**: $(n_x, n_y, n_{xy})$ E-PH pattern

Consider a victim cell in E which is interfered by adjacent aggressor cells. Let $n_x$, $n_y$ and $n_{xy}$ be the number of $x$-directional, $y$-directional and $xy$-directional (diagonal) aggressor cells in PH respectively. We will define this pattern as a $(n_x, n_y, n_{xy})$ E-PH pattern.

By Definition 1, we can reduce the types of E-PH patterns from $256 (= 2^8)$ into $45 (= |n_x| \cdot |n_y| \cdot |n_{xy}|)$ where $|n|$ is the cardinality of $n$ ($\because 0 \le n_x \le 2, 0 \le n_y \le 2, 0 \le n_{xy} \le 4$). In three dimensional memories such as vertical memories, each victim cell has 26 adjacent cells. In such three dimensional memories, the number of E-PH patterns will be reduced from $2^{26}$ to $30375$ ($= |n_x| \cdot |n_y| \cdot |n_z| \cdot |n_{xy}| \cdot |n_{yz}| \cdot |n_{zx}| \cdot |n_{xyz}|$) because of ($\because 0 \le n_x, n_y, n_z \le 2, 0 \le n_{xy}, n_{yz}, n_{zx} \le 4, 0 \le n_{xyz} \le 8$).

We can obtain $\max\{\Delta V_{(i,j)}\}$ and $\min\{\Delta V_{(i,j)}\}$ due to the cell-to-cell interference by (1). Unfortunately, a mathematical model for lateral charge spreading has not yet been developed. Therefore, we will only consider the cell-to-cell interference.

(2, 2, 4) E-PH pattern will cause the maximum threshold voltage shift of a victim cell $\Delta V_{(i,j)}$. From (1), $\max\{\Delta V_{(i,j)}\}$ is given by

$$\max\{\Delta V_{(i,j)}\} = (2\gamma_x^* + 2\gamma_y + 4\gamma_{xy}) \cdot \Delta V(\text{E} \to \text{PH}) \quad (3)$$

where $\gamma_x^*$ is used instead of $\gamma_x$ considering both indirect and direct cell-to-cell interference. $\Delta V(\text{E} \to \text{PH})$ is the threshold voltage change of the aggressor cell when it is programmed from E to PH. It is natural that $\Delta V_{(i,j)}$ will be zero when all aggressor cells are E.

Fig. 4 shows the distortion caused by the cell-to-cell interference in SLC flash memory. We assume that the threshold voltage distribution before the cell-to-cell interference is approximated by a Gaussian distribution $N(\mu, \sigma^2)$.

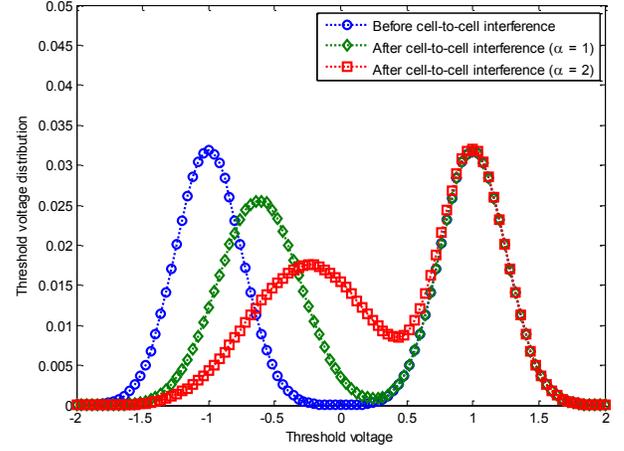

Fig. 4. Threshold voltage distribution before and after cell-to-cell interference. The threshold voltage distribution of $S_0$ before cell-to-cell interference follows $N(-1, 0.25^2)$ and that of $S_1$ follows $N(1, 0.25^2)$.

Based on [12], we set the coupling ratios $(\gamma_x, \gamma_y, \gamma_{xy}) = \alpha(0.1, 0.08, 0.006)$ where the scaling factor $\alpha$ is a parameter of cell-to-cell interference strength. The direct cell-to-cell interference was not considered in these coupling ratios. The scaling factor $\alpha$ and ratios between $\gamma_x$, $\gamma_y$ and $\gamma_{xy}$ are affected by device design. Though we assume that $\Delta V(\text{E} \to \text{PH}) = 2$, $\Delta V(\text{E} \to \text{PH})$ can be modified into random values. The result for random values will be presented in a future paper. Fig. 4 shows that the distortion caused by the cell-to-cell interference will be more serious as the scaling factor $\alpha$ increases.

III. MODULATION CODING SCHEMES FOR FLASH MEMORIES

In this section, we propose several modulation coding schemes for reducing the effect of E-PH patterns. In regard to the cell-to-cell interference of (1), our proposed schemes intend to reduce the number of aggressor cells whose threshold voltage shifts are large whereas other attempts focus on reducing coupling ratios or threshold voltage shifts of aggressor cells.

In this paper, we will focus on $x$-directional E-PH pattern which means reducing $n_x$. The reason is that $x$-directional E-PH patterns will be more critical due to direct cell-to-cell interference and lateral charge spreading as the cell size shrinks. In addition, program and read of flash memories are operated in a page unit which is a part of a word line. If any coding scheme is applied to several word lines, it will severely slow down the program and read speed performance. This is the reason why ECC is also applied within a page in most flash memory products. In order to reduce $y$-directional and $xy$-directional E-PH patterns by modulation coding schemes, we have to use two-dimensional coding schemes. However, two-dimensional coding will severely degrade the speed performance and be difficult to use in the current flash memory architectures. Previous research has also focused on one-dimensional coding schemes [13], [14]. Therefore, other signal processing solutions such as [12] will be more promising for reducing the effect of $y$-directional and $xy$-directional E-PH

TABLE II. BASIC ENCODING TABLE FOR (1, 7) RLL CODE

| Data | Codeword |
|---|---|
| 00 | 101 |
| 01 | 100 |
| 10 | 001 |
| 11 | 010 |

TABLE III. SUBSTITUTION TABLE FOR VIOLATIONS IN (1, 7) ENCODER

| Data | Codeword |
|---|---|
| 00 00 | 101 000 |
| 00 01 | 100 000 |
| 10 00 | 001 000 |
| 10 01 | 010 000 |

patterns.

### A. Run Length Limited (RLL) Codes

RLL codes have been applied successfully in both magnetic recording and optical recording. These codes have $(d, k)$ constraint to reduce the effects of intersymbol-interference (ISI) and prevent the loss of clock synchronization. $(d, k)$ RLL codes must satisfy the constraint that successive ones must be separated by at least $d$ and at most $k$ zeros [17], [18]. Using RLL codes, we propose novel modulation coding schemes for reducing $x$-directional E-PH patterns in SLC and MLC flash memories.

Tables II and III present a well-known (1, 7) RLL code. If no RLL violation would occur, the encoder uses Table II to encode the first pair. If an RLL violation would occur, the encoder encodes these two pairs using Table III [17]–[19].

### B. Modulation Coding Scheme for SLC

In SLC, there are only two states, namely, E ($S_0$: the erase state) and PH ($S_1$: the program state). If random data is programmed, $n_x$ can be 0, 1 and 2. By using the $d = 1$ constraint RLL code and NRZI coding, we can eliminate all $(n_x = 2, n_y, n_{xy})$ E-PH patterns.

**Example 1**. Original data are as follows.

| Index $i$ | 1 | 2 | 3 | 4 | 5 | 6 |
|---|---|---|---|---|---|---|
| Data | 0 | 1 | 0 | 0 | 1 | 0 |
| State | PH | E | PH | PH | E | PH |

In this example, data of '0' is mapped to PH and data of '1' is assigned to E, which follows the data mapping of most SLC flash memory products. $(n_x = 2, n_y, n_{xy})$ E-PH patterns are caused by the data pattern of '010' where the runlength of data '1' is one. In this example, there are two $(n_x = 2, n_y, n_{xy})$ E-PH patterns at $i = 1\sim 3$ and $i = 4\sim 6$. By using the $d = 1$ constraint RLL code and NRZI, we can make the minimum runlength of the same symbols into $2 (= d + 1)$. The following table shows that a (1, 7) RLL code of Table II, III and NRZI coding can eliminate all $(n_x = 2, n_y, n_{xy})$ E-PH patterns.

| Index $i$ | 1 | 2 | 3 | 4 | 5 | 6 | 7 | 8 | 9 |
|---|---|---|---|---|---|---|---|---|---|
| RLL | 1 | 0 | 0 | 1 | 0 | 1 | 0 | 0 | 1 |
| NRZI | 1 | 1 | 1 | 0 | 0 | 1 | 1 | 1 | 0 |
| State | E | E | E | PH | PH | E | E | E | PH |

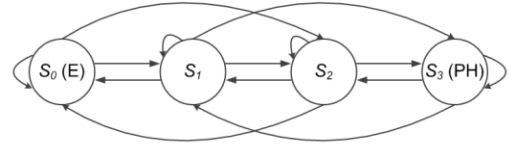

Fig. 5. State diagram withouth E-PH patterns for 2-bit/cell.

### C. Modulation Coding Scheme for 2-bit/cell MLC

#### 1) Binary Codes [16], [20]

The modulation coding scheme by using the binary RLL code is proposed to eliminate all $x$-directional E-PH patterns for 2-bit/cell MLC. From Table I, we know that consecutive ones in page 2 results in E-PH patterns since both E ($S_0$) and PH ($S_3$) have one in page 2. Therefore, we can prevent E-PH patterns by encoding only page 2 with the $d = 1$ constraint binary RLL code. Example 2 illustrates this scheme.

**Example 2**. Original data are as follows.

| Index $i$ | 1 | 2 | 3 | 4 | 5 | 6 |
|---|---|---|---|---|---|---|
| Page 1 | 1 | 1 | 0 | 1 | 0 | 0 |
| Page 2 | 1 | 1 | 1 | 1 | 1 | 0 |
| State | $S_0$(E) | $S_0$(E) | $S_3$(PH) | $S_0$(E) | $S_3$(PH) | $S_2$ |

In this example, a $(n_x = 1, n_y, n_{xy})$ E-PH pattern occurs at $i = 1\sim 3$ and a $(n_x = 2, n_y, n_{xy})$ E-PH pattern occurs at $i = 3\sim 5$. By using a (1, 7) RLL code, we will obtain the following coded data.

| Index $i$ | 1 | 2 | 3 | 4 | 5 | 6 | 7 | 8 | 9 |
|---|---|---|---|---|---|---|---|---|---|
| Page 1 | 1 | 1 | 0 | 1 | 0 | 0 | - | - | - |
| Page 2 | 0 | 1 | 0 | 0 | 1 | 0 | 1 | 0 | 0 |
| State | $S_1$ | $S_0$ | $S_2$ | $S_1$ | $S_3$ | $S_2$ | - | - | - |

In this coded data, all $x$-directional E-PH patterns ($n_x \geq 1$) have been eliminated, i.e., $n_x$ will be zero. Since only the data of page 2 has been encoded, we can store more data in page 1 than page 2. The states of $i = 7\sim 9$ depend on the additional page 1 data at $i = 7\sim 9$. Clearly, E-PH patterns will not occur at at $i = 7\sim 9$ because of $d = 1$ constraint in page 2.

The overall code rate of this coding scheme is $0.8333 = (1 + 2/3)/2$. The capacity will be $0.8471 = (1 + 0.6942)/2$ since the capacity of $d = 1$ constraint RLL codes is 0.6942.

#### 2) Quaternary Codes

The state diagram for eliminating all $x$-directional E-PH patterns for 2-bit/cell MLC is shown in Fig. 5 [16]. All paths between E and PH are excluded in this state diagram. The state transition matrix $T$ of the state diagram is given by

$$T = \begin{bmatrix} 1 & 1 & 1 & 0 \\ 1 & 1 & 1 & 1 \\ 1 & 1 & 1 & 1 \\ 0 & 1 & 1 & 1 \end{bmatrix}. \quad (4)$$

The capacity of this code is given by

$$C = \frac{\log_2 \lambda_{\max}}{M} \quad (5)$$

where $\lambda_{\max}$ is the largest eigenvalue of $T$ [18]. The capacity was divided by $M = 2$ for normalization. From (5), the capacity of the quaternary code is $0.9163 \, (\cong \log_2 3.5616/2)$. This is larger than 0.8471 which is the capacity of the modulation coding scheme using binary RLL codes.

TABLE IV. NOVEL QUATERNARY CODE FOR 2-BIT/CELL

| Subset index | Codeword candidates | The number of candidates without E-PH patterns within a codeword |
|---|---|---|
| 1 | 0xxx0 | 31 |
| 2 | 0xxx1 | 39 |
| 3 | 0xxx2 | 39 |
| 4 | 0xxx3 | 30 |
| 5 | 1xxx0 | 39 |
| 6 | 1xxx1 | 50 |
| 7 | 1xxx2 | 50 |
| 8 | 1xxx3 | 39 |
| 9 | 2xxx0 | 39 |
| 10 | 2xxx1 | 50 |
| 11 | 2xxx2 | 50 |
| 12 | 2xxx3 | 39 |
| 13 | 3xxx0 | 30 |
| 14 | 3xxx1 | 39 |
| 15 | 3xxx2 | 39 |
| 16 | 3xxx3 | 31 |
| Total number of candidates | | 634 |

'xxx' means any pattern without E-PH patterns

TABLE V. CODE RATES AND CAPACITIES OF PROPOSED MODULATION CODING SCHEMES WITHOUT $x$-DIRECTIONAL E-PH PATTERNS

| $M$ (bits/cell) | Binary RLL codes | | $2^M$-ary codes | |
|---|---|---|---|---|
| | Code rate | Capacity | Code rate | Capacity |
| 2 | 0.8333 | 0.8471 | 0.8000 | 0.9163 |
| 3 | 0.8889 | 0.8981 | 0.9333 | 0.9861 |
| 4 | 0.9167 | 0.9236 | - | 0.9973 |

Therefore, we can design quaternary RLL codes with higher rate than coding schemes using the binary RLL code. The quaternary codes will be illustrated in Example 3.

**Example 3.** The codebook for the novel quaternary code is presented in Table IV. In this codebook, we intend that each codeword candidate does not have any E-PH patterns. The total number of such candidates is 634. Though each candidate has no E-PH patterns, an E-PH pattern may occur in between a codeword candidate and the next one. By carefully avoiding E-PH patterns, we designed the following two codebooks.

*a) Codebook 1*

If we exclude subsets {1, 2, 3, 4, 5, 9, 13} having '0' in their first or last symbol, all $x$-directional E-PH patterns will be eliminated. The number of remaining candidates is 387 ($> 4^4$) and a codebook of rate 0.8 ($= 4/5$) can be constituted.

Instead of dropping subsets {1, 2, 3, 4, 5, 9, 13}, we can obtain a similar codebook without $x$-directional E-PH patterns by excluding subsets {4, 8, 12, 13, 14, 15, 16} having '3' in their first or last symbol.

*b) Codebook 2*

If we drop subsets {1, 4, 13, 16} that have '0' or '3' in both first and last symbol (i.e., the most undesirable subsets), the number of remainders will be 512. Thus, a codebook of rate 0.9 can be constituted and all $(n_x = 2, n_y, n_{xy})$ E-PH patterns will be eliminated.

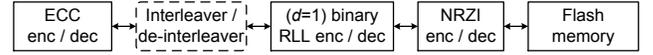

(a) Coding scheme for SLC.

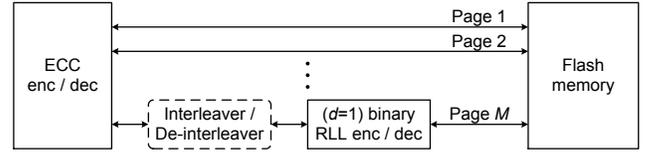

(b) Coding scheme with binary RLL codes for $M$-bit/cell MLC.

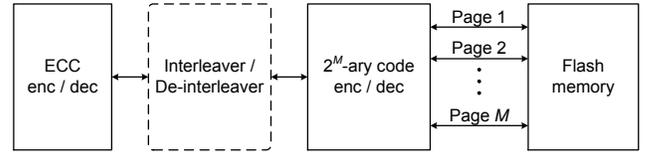

(c) Coding scheme with $2^M$-ary codes for $M$-bit/cell MLC.

Fig. 6. Block diagrams of proposed modulation coding schemes for SLC and $M$-bit/cell MLC.

Unfortunately, this codebook allows $(n_x = 1, n_y, n_{xy})$ E-PH patterns. These E-PH patterns can occur when a codeword of subsets {2, 3} is concatenated with that of subsets {8, 12}. In addition, a codeword of subsets {14, 15} and that of subsets {5, 9} can generate $(n_x = 1, n_y, n_{xy})$ E-PH patterns. If random data are programmed, the probability of $(n_x = 1, n_y, n_{xy})$ E-PH patterns will be 0.046 ($= 2 \times (78/512)^2$).

*D. Modulation Coding Scheme for M-bit/cell MLC*

For higher $M$-bit/cell flash memories such as 3-bit/cell and 4-bit/cell, modulation coding schemes can be designed in similar methods.

If the binary RLL code is used, the modulation coding scheme will depend on the mapping scheme that converts a state level to corresponding bit representation. In the widely used Gray mapping of flash memories, the last page bit of E ($S_0$) and that of PH ($S_{2^M-1}$) are generally one [21]. Therefore, all $x$-directional E-PH patterns can be eliminated by encoding the last page (page $M$) data with $d = 1$ constraint RLL codes.

$2^M$-ary codes can also be designed by similar methods with Example 3. For 3-bit/cell MLC, we designed three kinds of 8-ary modulation codes with code rate 8/9 ($\approx$0.8889), 11/12 ($\approx$0.9167) and 14/15 ($\approx$0.9333). The higher code rate requires longer codeword length, which results in a more complex codebook.

The code rate and the capacity of the scheme with binary RLL codes and the scheme with $2^M$-ary codes are presented in Table V. The capacity of the scheme with binary RLL codes is $\{(M - 1) + 0.6942\}/M$ and the capacity of the scheme with $2^M$-ary codes can be obtained using (5). Table V shows that we can design the coding scheme with higher code rate for larger bits per cell flash memory.

The block diagrams of the proposed modulation coding schemes are presented in Fig. 6. The interleaver can be included and it will be discussed in Section IV.

TABLE VI. SIMULATION PARAMETERS

|  | Conventional scheme | | Modulation coding scheme |
| --- | --- | --- | --- |
| ECC (BCH code) | $n$=4551, $k$=4096, $t$=35, $r$=9/10 | $n$=8191, $k$=4096, $t$=366, $r$=1/2 | $n$=5435, $k$=4096, $t$=105, $r$=3/4 |
| RLL code | - | - | (1, 7) RLL code (rate = 2/3) |
| Overall rate | 0.9 | 0.5 | 0.5 |

## IV. SIMULATION RESULTS

We will present simulation results of the modulation coding scheme for SLC. The reason is that the effect of E-PH patterns is most directly exposed in SLC because SLC has only E and PH. We assume that the number of ECC codewords per page is 16 (i.e., the information bits per page is 8KB). The threshold voltage distribution of $S_0$ before the cell-to-cell interference is assumed to $N(-1, 0.25^2)$ and that of $S_1$ to $N(1, 0.25^2)$ as in Fig. 4.

Table VI shows the simulation parameters. The conventional scheme is composed only a strong Bose-Chaudhuri-Hocquenghem (BCH) code. The proposed scheme is composed of a weaker BCH code, (1, 7) RLL code, NRZI, and interleaver. Parameters of BCH code are $n$ (codeword size), $k$ (information size), $t$ (error correcting capability), and $r$ (code rate). It is worth mentioning that $k$ of ECC is different from that of RLL codes.

In Fig. 7, we compare the threshold distributions of the conventional scheme and that of the modulation coding scheme. The modulation coding scheme can improve the threshold voltage distribution since all $(n_x = 2, n_y, n_{xy})$ E-P patterns are eliminated.

In Fig. 8, we compare the word error rate (WER) of the conventional scheme and that of the modulation coding scheme. WER is the probability of the ECC decoding failure. If $\gamma_x^*$ is small, the conventional scheme can be good enough to satisfy the reliability requirement. However, as $\gamma_x^*$ increases, the proposed scheme can cope with the cell-to-cell interference effectively. In regard to the decision level between $S_0$ and $S_1$, we used the scheme of [22] which sets the local minimum of the threshold voltage distribution between states as the decision level.

Whereas the interleaver does not improve the WER for the conventional scheme, the interleaver in the modulation coding scheme improves the WER significantly. The decoding failure of the RLL code results in error propagation and it means that burst errors can occur in one ECC codeword. The interleaver can spread these burst errors into other codewords and improve the WER of the modulation coding scheme.

If the interleaver size is larger than one page, it will slow down the program and read performance because the interleaver will need to wait to collect more than one page data. However, the interleaving within one page does not degrade the speed performance since one page is a unit of the program and read operations. In our simulation, the basic block interleaver [23] with size of only one page (i.e., 16 ECC codewords) has been used, which is shown in Fig. 9.

From the perspective of the hardware complexity, the modulation coding scheme will be much better than the conventional scheme. The BCH code of the modulation coding scheme will be much simpler than that of the conventional scheme because the codeword size $n$ is 66% of that of the conventional scheme and the error correcting capability $t$ is only 29%. The codeword size and the error correcting capability are pivotal parameters in hardware complexity of BCH codes. Since the hardware complexity of RLL coding and interleaving is small, the modulation coding scheme has advantages in not only reliability but also hardware implementation.

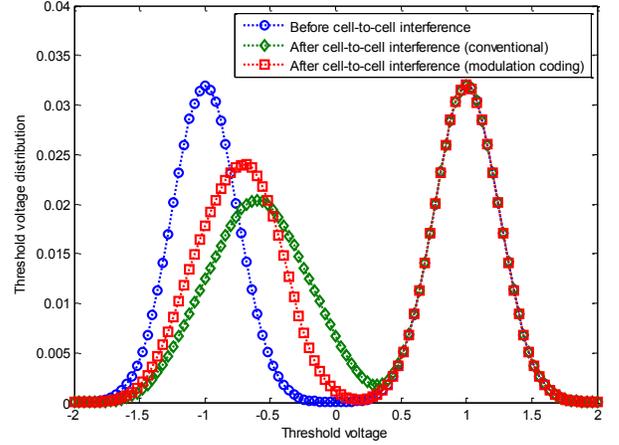

Fig. 7. Threshold voltage distribution of the conventional scheme and the modulation coding scheme $(\gamma_x^* = 0.2, \gamma_y = 0, \gamma_{xy} = 0)$.

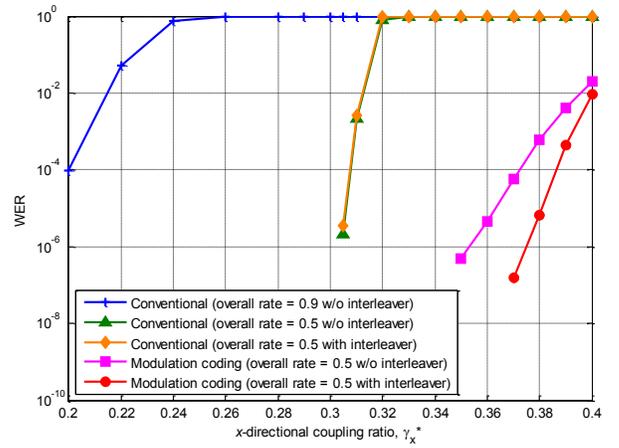

Fig. 8. WER of the conventional scheme and the modulation coding scheme.

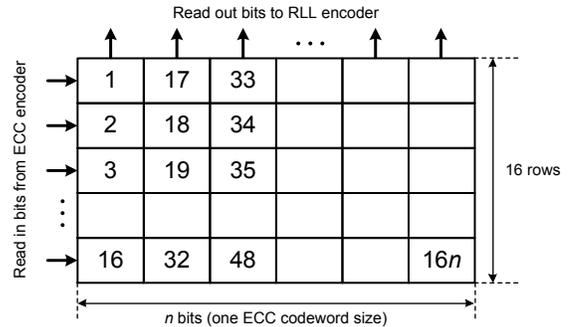

Fig. 9. Block interleaver with size of one page (i.e., 16 ECC codewords).

## V. CONCLUSION

In this paper, we defined E-PH patterns which degrade the reliability of flash memories due to cell-to-cell interference and lateral charge spreading. In order to cope with these degradations mechanisms, we proposed several modulation coding schemes which reduce the number of E-PH patterns. The modulation coding schemes can not only improve the reliability but also be beneficial for hardware implementation. These schemes would be more beneficial as the cell size scales down.

For future research, two-dimensional modulation coding schemes can be investigated in spite of the incompatibility with the current flash memory architecture.